\documentclass[a4paper,fleqn]{cas-sc}

\usepackage[authoryear,longnamesfirst]{natbib}

\usepackage{wrapfig}
\usepackage{lscape}
\usepackage{rotating}

\usepackage{cleveref}
\crefname{figure}{Figure}{Figures}
\crefname{section}{Section}{Sections}
\usepackage{graphicx}

\def\tsc#1{\csdef{#1}{\textsc{\lowercase{#1}}\xspace}}
\tsc{WGM}
\tsc{QE}

\begin{document}
\let\WriteBookmarks\relax
\def\floatpagepagefraction{1}
\def\textpagefraction{.001}

\shorttitle{Thematic Domain Analysis for Ocean Modeling}

\shortauthors{Reiner Jung, Sven Gundlach, and Wilhelm Hasselbring}

\title [mode = title]{Thematic Domain Analysis for Ocean Modeling}

\tnotemark[1]

\tnotetext[1]{This work is funded by the Deutsche Forschungsgemeinschaft (DFG, German Research Foundation), grant no. HA 2038/8-1 -- 425916241.}

\tnotetext[2]{This is a preprint of \url{https://doi.org/10.1016/j.envsoft.2022.105323}}

\author[1]{Reiner Jung}[type=author,orcid=0000-0002-5464-8561]

\ead{reiner.jung@email.uni-kiel.de}

\ead[url]{https://www.se.informatik.uni-kiel.de/en/team/reiner-jung}

\credit{Conceptualization, Implementation, Analysis, Publication}

\affiliation[1]{organization={Software Engineering Group, Department of Computer Science, Kiel University},
            addressline={Christian-Albrechts-Platz 4},
            city={Kiel},
            postcode={24118},
            country={Germany}}

\author[1]{Sven Gundlach}[type=author,orcid=0000-0003-4060-2754]


\ead{sven.gundlach@email.uni-kiel.de}

\ead[url]{https://www.se.informatik.uni-kiel.de/en/team/sven-gundlach}

\credit{Conceptualization, Implementation, Analysis, Publication}

\author[1]{Wilhelm Hasselbring}[type=author,orcid=0000-0001-6625-4335]

\ead{hasselbring@email.uni-kiel.de}

\ead[url]{https://www.se.informatik.uni-kiel.de/en/team/prof.-dr.-wilhelm-willi-hasselbring}

\credit{Conceptualization, Analysis, Publication}

\cormark[1]

\cortext[1]{Corresponding author}

\begin{abstract}
Ocean science is a discipline that employs ocean models as an essential research asset.
Such scientific modeling provides mathematical abstractions of real-world systems, e.g., the oceans. These models are then coded as implementations of the mathematical abstractions.
The developed software systems are called models of the real-world system.

To advance the state in engineering such ocean models, we intend to better understand how ocean models are developed and maintained in ocean science.
In this paper, we present the results of semi-structured interviews and the Thematic Analysis~(TA) of the interview results to analyze the domain of ocean modeling.
Thereby, we identified developer requirements and impediments to model development and evolution, and related themes.
This analysis can help to understand where methods from software engineering should be introduced and which challenges need to be addressed.

We suggest that other researchers extend and repeat our TA with model developers and research software engineers working in related domains to further advance our knowledge and skills in scientific modeling.

\end{abstract}


\maketitle

\section{Introduction}

In scientific modeling, mathematical abstractions of real world systems are drawn up and software is built as implementation of these mathematical abstractions.
The resulting software systems are called \textit{models} of the real world systems, which are then executed to simulate the real world.
These models are used to make predictions and to understand emerging properties of the real world that neither theory nor experiment alone is equipped to answer~\citep{Ruede2018}.

An application area of scientific modeling are ocean system models.
These ocean system models address certain aspects of the ocean, such as CO$_2$ uptake ~\citep{wanninkhof_global_2013}.
Many such models, which are complex software systems, have been developed and extended over decades by scientists from different disciplines~\citep{ClimateModels2015}.
Developing, extending, enhancing and integrating such models is a challenging endeavor and requires knowledge in multiple disciplines.
Thus, they are an interdisciplinary effort.

Ocean models have grown in size and complexity, which affects the effort necessary to improve and change models for addressing new research questions.
Despite the increasing importance of software-based modeling to the scientific discovery process~\citep{Ruede2018}, well-established software engineering practices are rarely adopted in computational science~\citep{CiSE2018}.

\citet{iwanaga2018software} report on software development practices in integrated environmental model development through literature review and expert knowledge such as code reviews, testing, and code management. They also emphasize the need for modular, component-based software structures for reusability and reproducibility in environmental model development. \citet{Iwanaga2021} further argue for strengthening interdisciplinary communication, and improvement of the softwre documentation process. They call for a grand vision for holistic system-of-systems research that engages researchers, stakeholders, and policy makers in a multi-tiered process for the co-creation of knowledge and solutions to major socio-environmental problems. With the present paper, we add a software engineering view to this interdisciplinary process of environmental model development.

To facilitate the transfer of knowledge in software engineering to computational science and scientific modeling, it is necessary for software engineers to understand the domain.
The best way to gain insights into a domain is observing and asking domain experts, in our case ocean model developers, model users, and research software engineers.
It is our main goal to understand how ocean modelers and engineers create, maintain, extend and use models.
Based on this understanding, we later intend to create tools and services that support the research work and improve collaboration in ocean modeling.
We chose semi-structured interviews as the method to address key issues, to understand how modeling is done in ocean science, and yet to be flexible enough to adapt if certain new aspects emerge.
We interviewed ocean model developers, scientific modelers and research software engineers who work in various research groups at universities and large-scale research facilities, such as the GEOMAR Helmholtz Centre for Ocean Research Kiel, the German Climate Computing Center (DKRZ) and the Max Planck Institute for Meteorology (MPI). 

For the analysis of the interviews, we chose the Thematic Analysis~(TA)~\citep{Braun2006} approach tailored to our questions.
Through the interviews with scientists from ocean modeling on the development and use of ocean models,
we learned that scientific modeling faces major challenges regarding understanding and reusing code, maintaining models, and introducing new scientists to scientific modeling.
Our employed research process may be repeated in related scientific disciplines.

The software and data availability is documented in Section~\ref{s-software}.
Our employed research methods are summarized in Section~\ref{sec-research-method}.
The results of the thematic domain analysis are presented in Section~\ref{sec-thematic-analysis}.
Section~\ref{sec-conclusion} draws our conclusions and indicates future research.

\section{Software and data availability}\label{s-software}

The complete thematic map as result of our thematic domain analysis is provided as supplementary material in the form of an online graph at:
\begin{quote}
	\url{https://oceandsl.uni-kiel.de/graph/}
\end{quote}
This online graph allows you to interactively explore our coding results.
In addition, the graph data and the visualization software is provided as Docker container open source at GitHub:
\begin{quote}
	\url{https://github.com/cau-se/oceandsl-interview-graph/}
\end{quote}
This enables you to install the online graph on your own hardware.

\section{Research Methods}\label{sec-research-method}

To study software development processes in ocean science, we conduct semi-structured interviews and analyze the interview results via the Thematic Analysis~(TA) approach~\citep{Braun2006}.
Section~\ref{sec-semi-structured-expert-interview} reports on our interview setup, before Section~\ref{ssec-thematic-analysis} explains how we used TA for analyzing the interview results.

\subsection{Semi-structured Expert Interviews}\label{sec-semi-structured-expert-interview}

As a qualitative research method, interviews allow to investigate new ideas, concepts and domains~\citep{seidman_interviewing_2019}.
Interviews can be characterized by the kind of interviewees (e.g., experts), number of interviewees, and how questions are asked ranging from only an initial question (narrative interview) to structured interviews with detailed questionnaires.
The semi-structured interviews, that were conducted in our study, lie in between narrative interviews and fully structured interviews.
We follow an interview guide which consists of introductory questions, e.g, profiles, followed by the main section with a few central questions and explanations.
The interview guide addresses the context and background of the interviewees, their work process, their tool environment, and whether they use software engineering methods.
The questions and explanations are used to steer the interviews, but still allow enough room to explore the overall topics of the interviews.
Due to the SARS-CoV-2 pandemic, most of the interviews were conducted via video conference. 

\paragraph{Employed Software Tools}
Thematic Analysis~(TA) requires complete transcriptions of the interviews.
All of our interviews were recorded in an audio format for later automated speech recognition.
As speech recognition software, a Kaldi Gstreamer Server was used with the latest pretrained model~\citep{milde-koehn-18-german-asr}.
The finalization was done by manually proofreading the transcript and analyzing the raw data using the audio software ocenaudio~\citep{ocenaudio}.
For the further TA, including familiarization, generation, searching and reviewing, the R package for Qualitative Data Analysis RDQA was employed~\citep{RQDA}.
Final processing, including defining and producing codes, was done using a SQL database, customized R-scripts and Java programs.

\paragraph{Participants}
We interviewed ten interviewees, always with two interviewers.

\paragraph{Structured Interview Guide}
The interviews were structured into introduction, process, environment and methods, and were directed toward the modeling process.
Each part was used for an initial theme of the TA, namely Interviewee Profile, Software Development Processes, Infrastructure Environment and Software Engineering Methods:
\begin{itemize}
  \item \textbf{Interviewee Profile} The first introductory questions address the background of the scientists and engineers as well as their job characteristics and research interests. This way, we identify the demographics of our participants. This allows to understand the involved disciplines and topics addressed within the domain.
  \item \textbf{Software Development Processes} The second part is directed on how scientists and engineers work in ocean modeling.
Here, the main interest is in their work processes and interactions with others.
  \item \textbf{Infrastructure Environment} The next block addresses the software and infrastructure environment used to develop, test, and run their models.
This is important, as the acceptance of new tools and practices depends on the ability to adopt them with reasonable effort.
  \item \textbf{Software Engineering Methods} The interview closes with questions regarding software engineering methods, such as testing, static code checking, and feature management.
\end{itemize}

\subsection{Thematic Analysis of the Interview Results}\label{ssec-thematic-analysis}

Qualitative research provides a wide area of different methods and methodologies to analyze data and especially interviews.
These methods are usually developed in the social sciences and then transferred to other disciplines, including software engineering. 
For this study, we chose Thematic Analysis~(TA) as the method to analyze our interview data. 
TA has a long history which is widely used in psychology and the social sciences to analyze qualitative data including  images, videos and interviews.
Following the conceptualization of \citet{Braun2006} it is a flexible method that can be customized towards our research goals.
The central concept of TA are themes.
A \emph{theme} captures an idea, concept or pattern within the data which seems relevant in relation to the research question.
The relevance of a theme is not defined by quantifiable measures, but qualitatively by the \emph{relevance} with respect to the research goals.

TA can be used to provide a \emph{descriptive} result for the whole data set, e.g., all interview transcripts, or for an in-depth analysis to identify specific latent themes.
A \emph{latent theme} in TA is a theme which is not explicitly mentioned in the questions and answers during an interview.
Instead, it describes an underlying pattern.
For example, the pattern that there are certain roles in the ocean modeling domain and that there are competing priorities between these roles.
In contrast, a \emph{semantic theme} is closer to the data set and more descriptive.
TA can be applied to any media, including interviews and transcripts.
All media are considered \emph{data} and all media used for a specific TA is called a \emph{data set}.

In our study, we combined inductive and deductive analysis.
\emph{Inductive analysis} is a bottom up approach where coding is done without a pre-existing coding frame or an analytic preconception.
In an inductive analysis, the underlying paradigm is constructivist and focuses on socio-cultural contexts and structural conditions.
Opinions and experiences of interviewees are seen as socially (re-)produced entities.
This method is mainly useful when analyzing a previously unknown domain, or to avoid influencing the analysis results by personal bias.
\emph{Deductive analysis}, in contrast, is driven by a particular research interest in an area.
Thus, codes and themes are influenced by research questions.
Essentially, an inductive analysis allows to identify properties of a domain that were not previously considered.
This is especially useful when trying to understand new domains.
However, it may produce less specific answers regarding a specific research goal.
Whereas a deductive analysis narrows the focus to specific research interests and provides more in-depth results, it may miss properties that were not considered in advance.
We started with specific research questions and, therefore, with a deductive analysis but complemented this by a subsequent inductive analysis.

\paragraph{Qualitative data analysis}
Qualitative data analysis examines and interprets data to understand what it represents.
The TA method comprises six phases which can be re-iterated multiple times, depending on modifications to the themes, codes and text extracts:

\textbf{Phase 1: Familiarization with the data\ }
The researcher familiarizes him- or herself by actively reading the whole data set -- at least once -- before coding.
\textit{Coding} is the process of labeling and organizing qualitative data to identify different themes and the relationships between them. 
During familiarization, the researcher collects initial ideas and patterns, including the reasons for their relevance.

\textbf{Phase 2: Generate initial codes\ }
Based on these ideas, the researchers identify an initial set of codes.
When coding, they assign labels to words or phrases that represent important (and recurring) themes in each response. 
In a data-centric approach the codes rely primarily on the data, while in a theory-centric approach, the codes are influenced by the set of research questions the researchers have towards the data set.
Each code refers to a meaning within a data extract.
However, a text extract may relate to multiple codes, such as the text extract \emph{`we use git to manage code and contributions'} may be coded as both \emph{`Version control with git'} and \emph{`Code management'}.
Thus, this phase produces a set of codes, where each code is associated with one or more extracts from the whole data set.

\textbf{Phase 3: Searching for themes\ }
Themes represent patterns and relevant topics -- also referred to as narratives -- occurring in or emerging from the interviews.
To create themes, codes are grouped and subsequently associated with potential themes.
Themes can be interrelated.
The relationship includes contrasting views and hierarchy, i.e., a theme to sub-theme relationship.
This results in a \emph{theme map} that contains a set of candidate themes, related to codes and data extracts.
In this phase, no code or extract gets discarded.
In case a code does not (yet) match any theme, it can be associated to a theme called \textit{miscellaneous}.

\textbf{Phase 4: Reviewing and refining themes\ }
The review and refinement of the thematic map comprises two steps.
First, each theme is analyzed regarding its associated extracts to check whether they form a coherent pattern, i.e., they address the same issue, topic or narrative.
In case an extract does not fit, it is discarded or moved to another theme.
Here, themes can be created, split, merged or removed.
Second, the resulting themes must be reflected based on the whole data set, if they are accurate representations of the data set.
Therefore, it is crucial to revisit the associated text segments and their context. 
In case relevant aspects are missing, additional codes can be added.

\textbf{Phase 5: Defining and naming themes\ }
The essence of each theme is identified and an internally consistent account of the theme is compiled.
In case themes are too complex and diverse, they are split and further categories are identified.

\textbf{Phase 6: Produce the report\ }
Finally, a report is compiled covering all themes.
The report must cover the complete story of the whole data, provide specific descriptions of all themes and their relationships.

\begin{figure}[!htb]
	\centering
		\includegraphics[width=\textwidth]{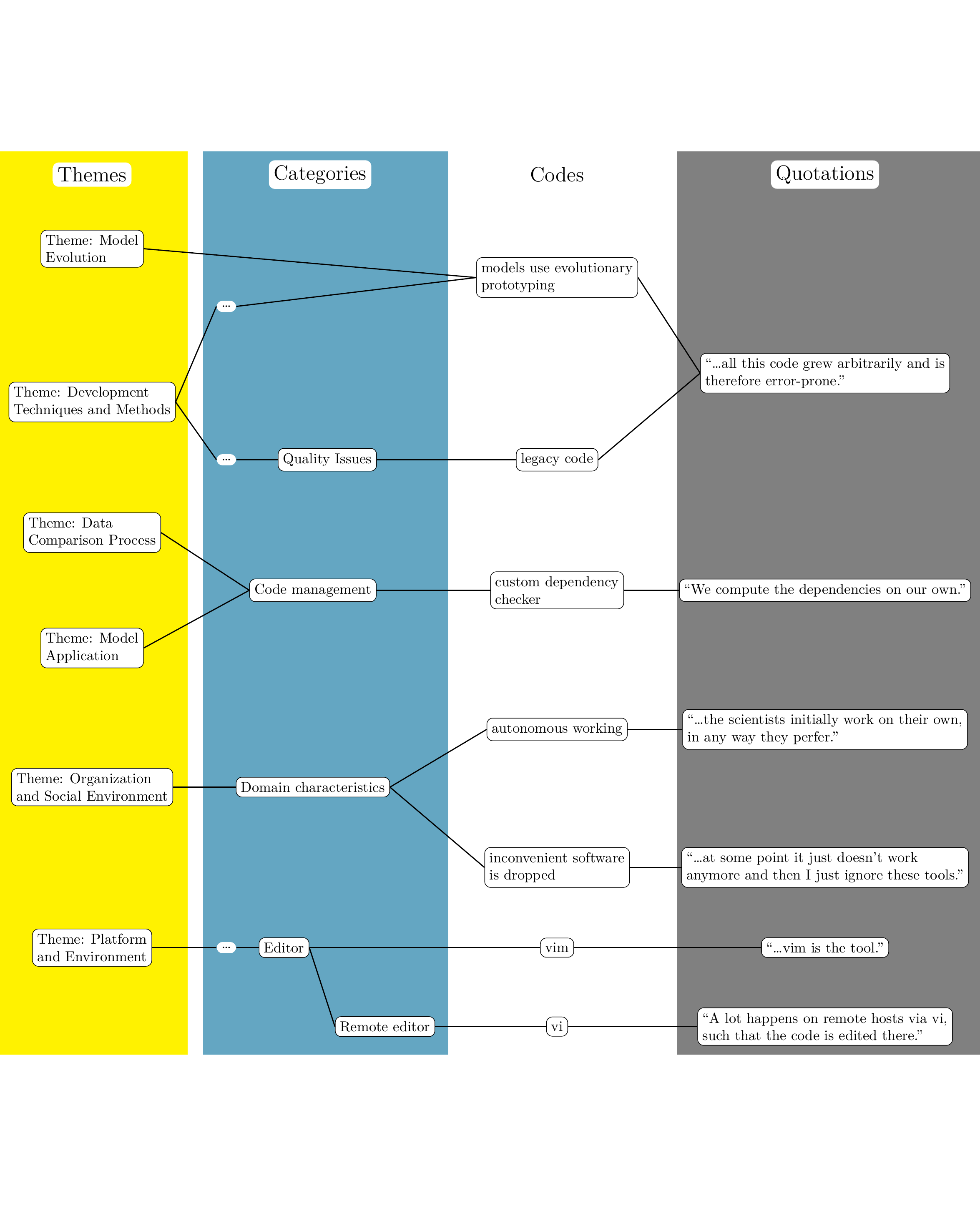}
	  \caption{Example TA Transformation process of quotations from domain experts that are transferred into codes, hierarchically categorized, and assigned to themes; from right to left.}\label{ta-quotes-table}
\end{figure}

\textbf{Example\ } \cref{ta-quotes-table} illustrates the TA process transforming quotations from domain experts into codes, that are further categorized into sub-categories and associated to themes. The example quote ``\dots all this code grew arbitrarily and is therefore error-prone'' is coded as ``models use evolutionary prototyping'' for the theme Model Evolution and ``legacy code'' in the sub-category ``Quality Issues'' for the theme ``Development Techniques and Methods'' (from right to left in \cref{ta-quotes-table}).

\textbf{Comment\ } Please note that in the context of the present paper, the terms code and coding are used in two contexts: coding of interview data with TA and coding / programming of ocean models via some programming language.
The TA code ``legacy code'' from the previous example refers to outdated program code as part of ocean models.

\section{Thematic Domain Analysis Results}\label{sec-thematic-analysis}

Based on the Thematic Analysis~(TA), we gained insight into the work processes and roles of scientists and engineers in ocean modeling.
We analyzed the interviews utilizing TA as described in Section~\ref{ssec-thematic-analysis} to identify multiple themes that support a general understanding of the way scientists and engineers work in the domain.
We followed the six phase process of TA, as introduced in Section~\ref{ssec-thematic-analysis}.
For the transcription, we relied on speech to text software to support the text generation (cf. Section~\ref{sec-semi-structured-expert-interview}).
The result required intensive manual adjustments due to the specific topic of our study.
We were able to familiarize ourselves with the interview text and audio in great detail in this phase.

Based on the transcripts, we identified initial code candidates and started coding with the RQDA coding tool~\citep{RQDA}.
We created codes based on the interview data, following a data-centric TA approach, and selected codes and categories.
During the initial coding, we generated new codes for each interview resulting in some duplication, which had to be merged at a later stage.
Based on the coding, we identified processes and role-related themes, as well as other domain-specific themes.
We discovered relationships between these themes and mapped all codes to at least one theme.
Subsequently, we reviewed and refined the identified themes following the approach in Section~\ref{ssec-thematic-analysis}.
Thus, similar codes were merged, codes were moved to more fitting themes and removed if they were out of scope.
Finally, all interviews were re-read, reflecting each theme.
During this process, new codes and themes emerged, which were addressed accordingly.

\begin{figure}
	\centering
	\includegraphics[width=\textwidth]{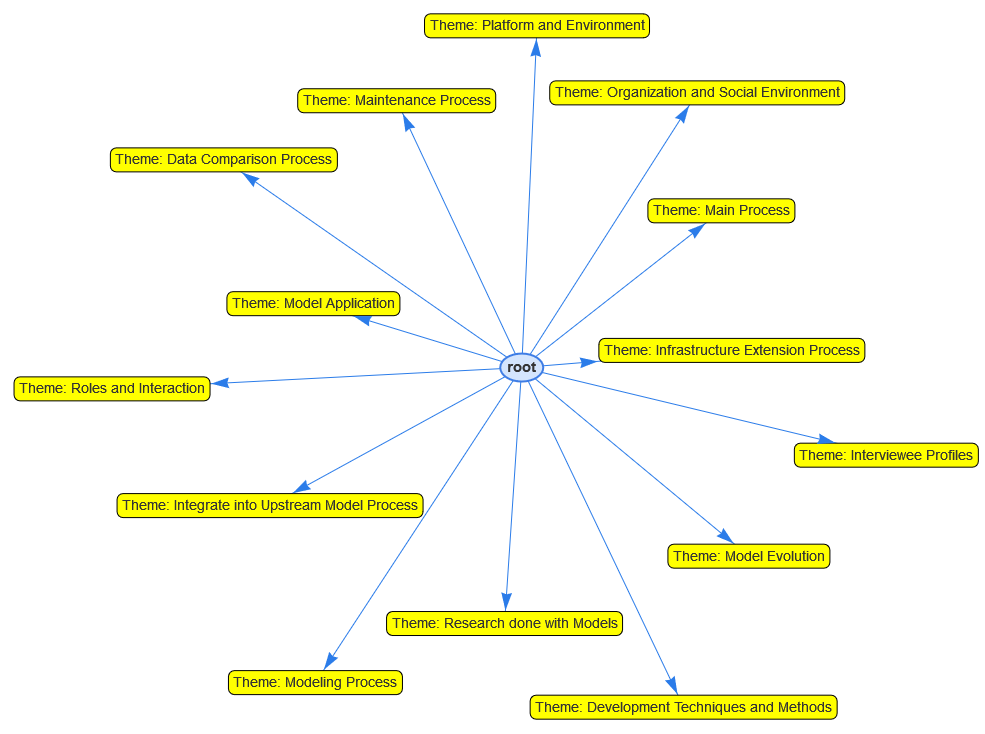}
	\caption{Top-level view of the thematic map depicting the resulting themes (yellow). Refer to \url{https://oceandsl.uni-kiel.de/graph/} for the complete, interactive map.}
	\label{fig-thematic-map-root}
\end{figure}

As a result, we obtained a large theme map.
We identified fourteen top-level themes in the interviews, as shown in \cref{fig-thematic-map-root}.
The complete map is provided as supplementary material in the form of an online graph at
\begin{quote}
	\url{https://oceandsl.uni-kiel.de/graph/}
\end{quote}
This online graph allows you to interactively explore our coding results.
In the present paper, we only display an extract, which is relevant for this paper.

In the following subsections, we will present the interviewee profiles (\cref{sec-interviewee-profiles}), the platform and environment the scientists and engineers work on (\cref{sec-platform-end-environment}), the simulation execution (\cref{ss-simulation}), the evolution of model code (\cref{ss-model-evolution}), the development techniques and methods (\cref{sec-development-techniques-and-methods}), and the goals of working with models (\cref{ss-research-done}) that were derived from this thematic map. The identified roles and development processes are presented separately by~\citet{ProcessesArXiv2021}.

\subsection{Interviewee Profiles}
\label{sec-interviewee-profiles}

\begin{figure}
	\centering
	\includegraphics[width=.8\textwidth]{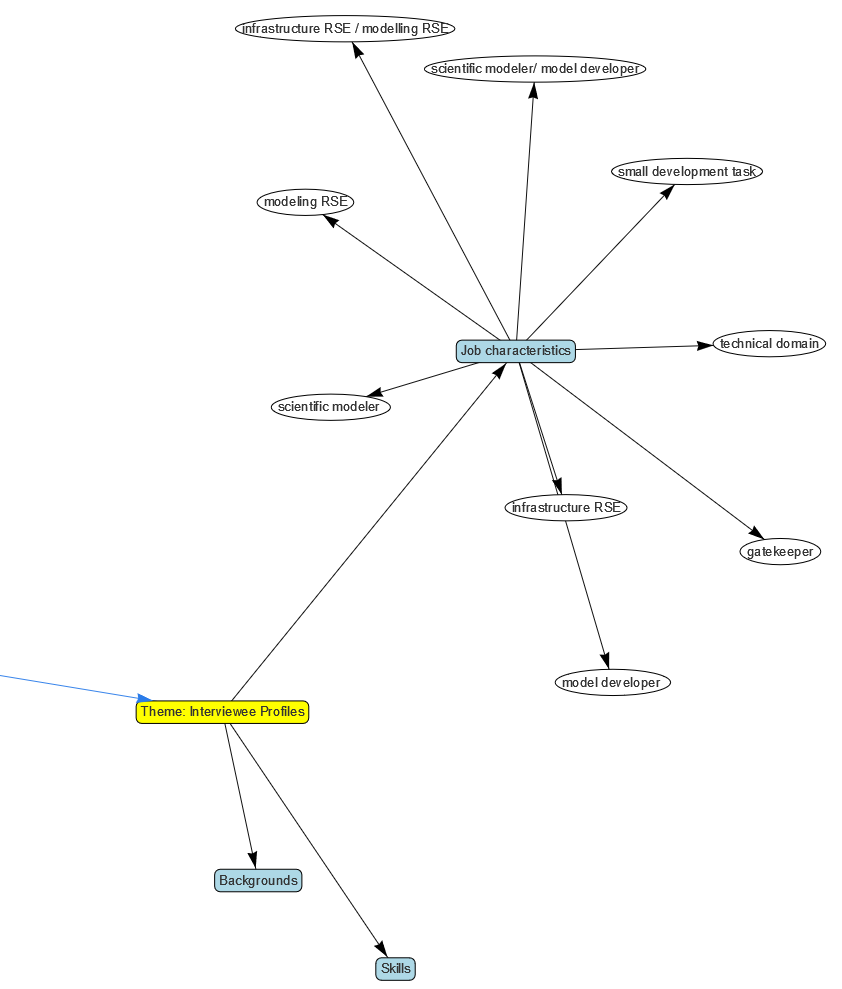}
	\caption{Excerpt of the unfolded thematic map depicting the skills and job characteristics for the interviewee profiles. The top-level themes from \cref{fig-thematic-map-root} are displayed in yellow (only Interviewee Profiles in this extract), the sub-categories in blue and the codes in white. We use the same color scheme as was used in \autoref{ta-quotes-table} on Page~\pageref{ta-quotes-table}.}
	\label{fig-ThemeIntervieweeProfiles}
\end{figure}

We held our interviews with domain scientists from Kiel University, the GEOMAR Helmholtz Centre for Ocean Research Kiel, the German Climate Computing Center (DKRZ) and the Max Planck Institute for Meteorology (MPI).
Their background is from various scientific disciplines including mathematics, oceanography, physics, meteorology, and biogeochemistry.
\cref{fig-ThemeIntervieweeProfiles} displays an excerpt of the partially unfolded thematic map depicting the skills and job characteristics for the interviewee profiles (for a closer look, we suggest the interactive online version).

The scientists work as model developers for their specific area, but are model users regarding other areas, e.g., they develop ocean circulation models, but use an atmosphere model. 
Thus, they often take multiple roles. 
Some scientists are now research software engineers and support other modelers with their modeling tasks.
They coordinate joint efforts in projects and act as consultants and trainers for scientists.

Scientific modelers add and modify the code to integrate new ocean processes and add new model components.
They perform model coding with support from research software engineers where needed.
Some research software engineers focus on optimizations for (a) the parametrization and (b) the parallel execution performance of models.

Scientists are also involved when adapting models to new time discretizations and spatial resolutions.
They work on scenario building, which also addresses resolution, selection of processes and model components.
In this area, they may coordinate their efforts with research software engineers or rely on their support.
The latter provide support in the development, installation and adaption of models to specific hardware.
Thus, research software engineers play an integral part in model migration.
They advise scientists in coding and development, and perform maintenance tasks, such as code cleanups and performance optimizations.

The software environment of the interviewees regarding modeling is dominated by long-living software systems \citep{Goltz2015}, which have grown over decades (20--30 years).
The model code is large, often written in various Fortran dialects.
To contribute to these models, scientists have to gain access to the whole code and understand the scientific processes represented in the model.
Code comprehension is also required to debug and test any contribution, as the architecture of these models is not fully documented.

\subsection{Platform and Environment}
\label{sec-platform-end-environment}

Here, we describe the development and runtime platform used for modeling and the tools used in modeling and model maintenance.
In ocean modeling, there exists no real separation between development and runtime platforms.
These platforms are diverse and depend on the employed community models.
Larger models are often developed, tested and maintained directly on the remote (HPC) platform they run on.
This is necessary, as there are often platform-related dependencies.
Therefore, we describe in the following the diversity of the both development and runtime platforms together starting with the hardware and operating systems,
followed by programming languages and tooling.

The scientists and developers use workstations PCs with various operating systems, computing clusters and HPC infrastructure.
Most hardware platforms run some Linux distribution and some workstations run macOS.
The HPC systems provide queuing systems such as Slurm to queue and execute model runs and data analysis.

Models are predominately written in Fortran.
Some models also use C and C++ code.
Matlab and Octave are used for smaller models, prototyping and data analysis.
R and increasingly Python are also used in post-processing and data analysis.

Our interviewees do not use integrated development environments~(IDEs), such as Eclipse, for their development. These IDEs are not available on all platforms and appear too cumbersome to run.
However, some use PyCharm for local development of Python code.
Most coding is done with plain text editors, where vi and derivatives are in the lead, especially when model developers and users switch between remote and local environments.
On macOS, scientists preferred Xcode for local development.
Furthermore, some scientists use Emacs and NEdit for their work.

\begin{figure}[!htb]
	\centering
	\includegraphics[width=\textwidth]{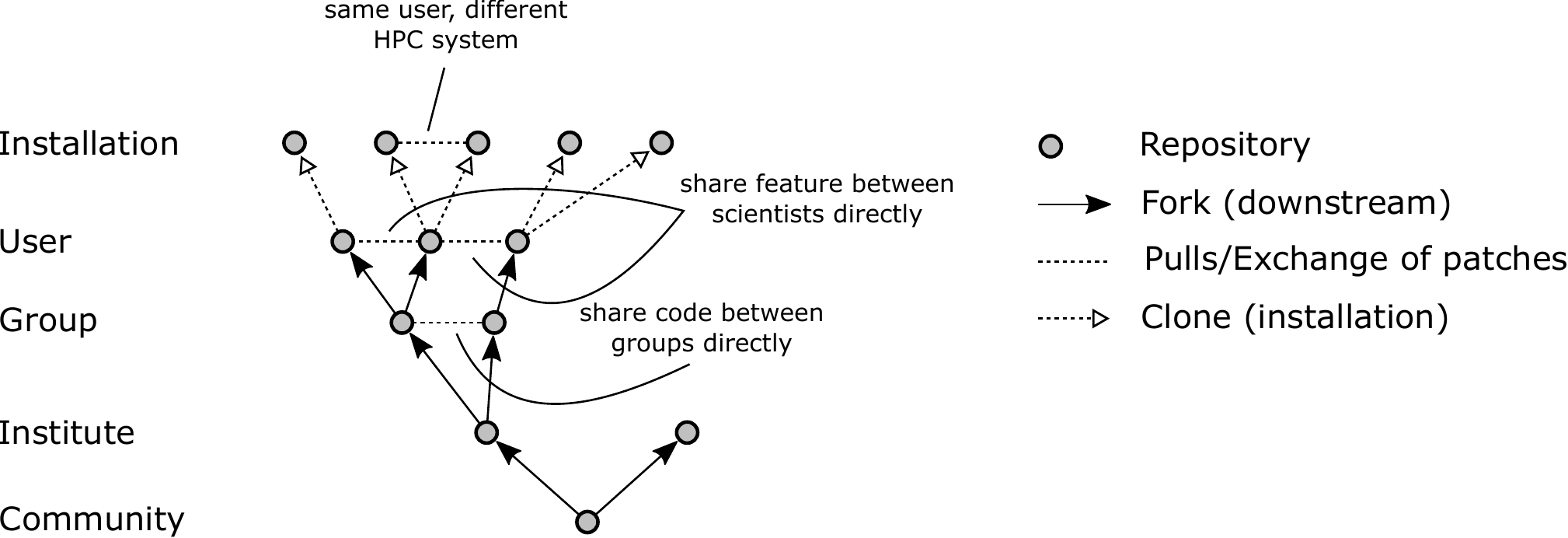}
	\caption{Example illustration for the relationships of repositories and the exchange of contributions in a modeling project}
	\label{fig-example-repository-relationships}
\end{figure}

Code variant and version management is done with Git, combined with Rsync for fast incremental file transfer.
In some cases, research software engineers use tarballs, but within the group of our interviewees, Git is the preferred tool to manage code.
There can be five different layers of Git repositories, which also use forks and feature branches (see \cref{fig-example-repository-relationships}).
Not all types of repositories and their relationships are used in all projects.
The main upstream model is the community model (bottom of \cref{fig-example-repository-relationships}).
Derived from the community model, each institute has an institutional repository.
In our interviews, these institutional repositories all used Git.
Beneath the institutional repository, there are either research-group-specific repositories or individual user repositories.
The lowest two levels are user-level repositories.
Usually, scientists have a repository on different HPC installations, managing configurations and hardware adaptations by branching.
Extensions and improvements realized on user level may be passed upstream to the research group or institutional repository when considered relevant for the
future model development in the research group or institute.
In the same way, the contributions can be submitted to the community model itself.
However, it is also possible to exchange contributions directly between modelers via Git pulls.

Builds are automated via tools like Autoconf and Make, or project-specific setup scripts programmed in shell languages.
However, there are some activities to use GitLab-CI as a service to control and execute builds.
The execution of models is often done on the command line utilizing a queuing system.
For post-processing, some interviewees use Jupyter notebooks.
These notebooks are also used to provide some documentation.

Specific for ocean models are tools to couple and nest models.
The NEMO model, for instance, uses AGRIF \citep{AGRIF2016} to nest models.
The coupler OASIS3-MCT \citep{OASIS3-MCT2017} can be used to couple a wide range of different models.
Dependency management is often done in this context by self-written tools.

To summarize, all the tooling in this context is heterogeneous and lacks standardization.
\autoref{tbl1} provides an overview of selected tooling employed in the context of our interviewees.

\begin{table}[htb]
\caption{Selected Tooling used by our Interviewees}\label{tbl1}
\begin{tabular}{l|l|p{8cm}}
	\toprule
	\textbf{Category} & \textbf{Tool} & \textbf{Notes and Usage} \\
	\midrule
	\multicolumn{3}{l}{Editing} \\
	\hline
	& vi/vim & Most used editor, used locally and remote \\
	& Emacs/Aquamacs & Used as editor supporting IDE-like features \\
	& XCode & Primary editor on macOS \\
	& PyCharm & Used in some cases for Python development \\
	\hline
	\multicolumn{3}{l}{Compilation} \\
	\hline
	& C, C++, Fortran & Host specific compilers from Intel in various versions \\
	\hline
	\multicolumn{3}{l}{Code Management} \\
	\hline
	& git & Managing versions and variants of model code \\
	& rsync & Fast incremental file transfer \\
	& tarballs & Legacy approach to share code \\
	\hline
	\multicolumn{3}{l}{Testing} \\
	\hline
    & Testing frameworks & Self-build frameworks sued with specific models and institutions \\
  	& MatLab/Octave & Testing models for plausibility \\
    & GitLab CI & Used to control the build of models (experimental stage) \\
	\hline
	\multicolumn{3}{l}{Build Tools} \\
	\hline
   & Make &  Central build tool used in most projects \\
   & Automake & Automatic generation of build scripts \\
	 & Autoconf & Automatic configuration tool\\
   & Conda & Python package tool \\
	\hline
	\multicolumn{3}{l}{Model Couplers} \\
	\hline
	& AGRIF & Coupling models \\
	& OASIS3-MCT & Coupling models \\
	\hline
	\multicolumn{3}{l}{Data Analysis} \\
	\hline
	& R & Statistical analysis \\
	& Python & Data analysis \\
	\hline
	\multicolumn{3}{l}{Project Management} \\
	\hline
    & Kanban board & Managing tasks \\
	\bottomrule
\end{tabular}
\end{table}

\subsection{Executing and Simulating Models}\label{ss-simulation}

The execution and simulation of models by model users and developers varies on how much care the model requires during the run and the different phases of a run, as well as, whether these executions are test runs and plausibility checks.
While some aspects are similar across all models and model setups, there are also specific differences induced by the models and couplers.

Our interviewees use the Earth system climate model UVic \citep{UVic2001}, the atmosphere model ECCHAM in different versions \citep{ECHAM2013}, the ocean model NEMO 3 \citep{NEMO2015}, the Earth system model ICON \citep{ICON2020}, and a marine ecosystem model toolkit Metos3D \citep{Piwonski2016}.
All these models consist of other models and modeling components that are coupled via couplers such as OASIS3-MCT \citep{OASIS3-MCT2017} and AGRIF \citep{AGRIF2016}.
All these parts require specific configurations and attendance.

Beside configuration and runtime monitoring, data management is a critical aspect of model execution.
At startup, models require parameters, data to configure the environment, and model state information collected from previous runs.
The data can have different versions and variants depending on their source and their customization for a specific experiment.

Input data is stored in Git repositories and specific data repositories.
In some modeling setups, the workflow executing a model experiment is able to fetch input data dependencies automatically via their Git revision tags.
This procedure is employed when models are used together with observation data, which can change regularly.
Output data is usually stored on the HPC system, as they can be huge.
Furthermore, the output is often stored also in Git repositories alongside the configuration and code of the model, as they all belong together.
Most modeling projects use the NetCDF format and special libraries to store input and output data.
NetCDF can contain metadata describing the content and the format.
NetCDF is a binary format and requires special tools.
\subsection{Model Evolution}\label{ss-model-evolution}

Model evolution addresses the continual changes applied to the model's code and documentation. 
Model code has grown into a complex structure created by many different developers with different skill sets and level of expertise.
Like in many long-living and large projects \citep{Goltz2015,reussner_managed_2019} no single person is able to understand all parts of the code.
In combination with a non-modular architecture, tight coupling and technical debt due to unstructured legacy code, changes are expensive.
\cref{fig-ThemeIntervieweeProfiles} displays an excerpt of the unfolded thematic map for the theme Model Evolution.

\begin{figure}[!htb]
	\centering
	\includegraphics[width=\textwidth]{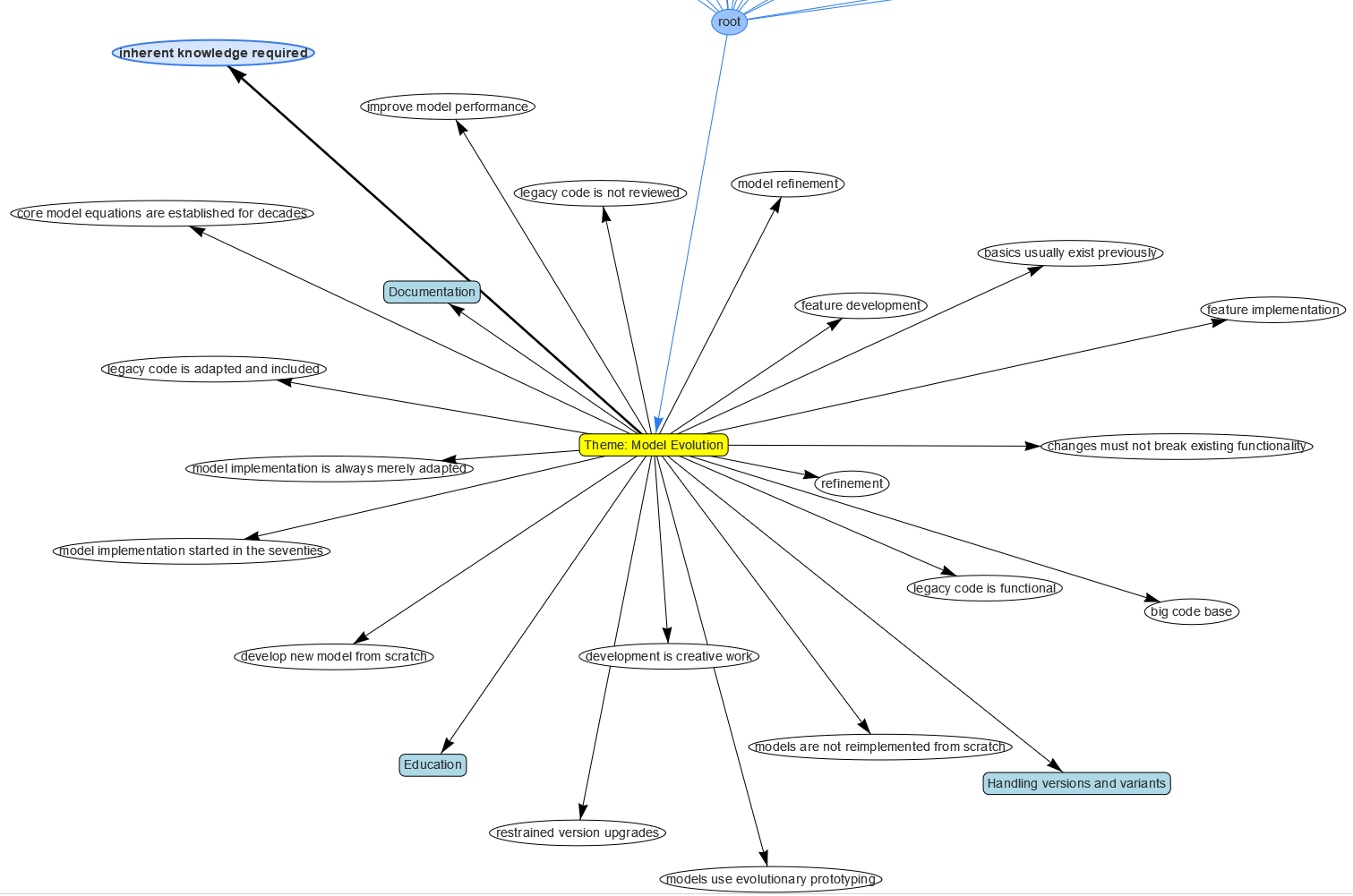}
	\caption{Extract of the unfolded thematic map for the theme Model Evolution (yellow). Again, the sub-categories are displayed in blue and the codes in white.}
	\label{fig-ThemeModelEvolution}
\end{figure}

Research software engineers emphasize that the code quality suffers specifically due to inappropriate compiler settings which
ignore warnings, lack of testing, short term focus on features instead of code quality, and lack of training in software engineering skills.   

The issues around code quality are complemented by issues in the documentation (category Documentation in \cref{fig-ThemeModelEvolution}).
The technical documentation for architecture, dependencies, features, and APIs is mostly missing, incomplete, outdated, contradictory, 
or wrong.
For ocean models, program documentation is rarely available.
As mathematical descriptions in papers are often on a different level of abstraction than the implementation, a synchronization between research paper and code is not easily achievable.

Dependency resolvers and build tools widely used in other areas of software development tend to be inappropriate in the modeling domain, due to conditional compilation dependency branches which are managed via preprocessor marks.
This results in self-built tools to identify and resolve such dependencies.

The issues with the code, documentation and tooling have impact on the way research software engineers work with models.
Model developers require access to the complete code of a model to understand it and to review code contributions from other developers.
This leads to a high implementation effort.
The lack of a modular architecture prevents the application of modern testing concepts.
Which also leads to a high implementation effort.
For example, in one reported case, the model development took one year, while implementation, debugging and testing required an additional three years. 

Research software engineers and model developers pointed out that many of the issues have a project management component to it.
Missing rules and the lack of enforcement of rules have an impact on the code quality.
This is also caused by limited accountability for contributions by their respective contributors.
In the past, contributors were not instructed or encouraged to follow coding standards and guidelines, which also still results in individual coding styles, as this is still the cases in many settings reported cases.
According to the interviews, this is also as there are no agreed governance strategies implemented to ensure quality measures.
Scientists are usually discouraged to follow coding and documentation guidelines, as this requires additional time and effort on their part.
Time they could spend writing papers and performing research.
As the \emph{currency} in science are publications, any measure, which reduces the time available for publications, will most likely not be successful without other incentives.

Finally, there is usually no overall strategy for code deprecation. 
Thus, all contributions stay in the code and hinder the evolution.
This is critical as models are often tailored to specific hardware installations, i.e., one specific HPC machine utilizing one specific compiler.
In case this HPC is retired, the code must be migrated to a new platform, regardless if parts are deprecated and no longer used. 

\subsection{Development Techniques and Methods}
\label{sec-development-techniques-and-methods}

This section discusses two essential development methods and techniques used in
modeling: testing and modular software architecture.

\paragraph{Testing} Testing is a central element of quality assurance in software engineering.
It is usually based on a large set of unit tests.
On top of this, integration tests evaluate whether units work in collaboration with other units. 
However, in scientific computing, the identification of small enough units to test is challenging.
Model components are interdependent with other components and their computation results are not known upfront.
Thus, testing is done either with complete reference experiments where result data is present which is then used to check the new version against the old results (regression tests), or the results are scrutinized by scientists whether the results are plausible.
The usually unknown output of the complete ocean model is known as the \textit{test oracle} problem in software engineering~\citep{Oracle2013}.
However, some interviewees assumed that smaller units in models would possibly make unit tests feasible.
In the area of infrastructure code, like logging and IO, unit testing is applied by some modeling projects.
Unfortunately, even in newer modeling projects, increasing test coverage has not been achieved.
Interviewees suppose that this originates from two influence factors:
\begin{itemize}
	\item New modelers are not required to add tests for their contributions, and no additional incentives are given.
	\item Current test facilities and practices for model code are not mature enough to establish an approach to implement tests.
\end{itemize}
While some projects encourage testing, interviewees suggested that without incentives the tests will not be implemented, as scientists rather use their time for publications. 

Despite the limits in tooling, techniques and methods, scientists and engineers aim to improve coding quality by experimenting with new testing concepts. They are open-minded for new tools and methods.
Testing is considered relevant for the community, and they are willing to try new concepts to improve quality.
Furthermore, research software engineers expressed interest in tools to statically check code quality automatically, but were unfamiliar with the concept and did not know tools able to perform such tasks, especially for Fortran, C, and C++.

\paragraph{Modular Software Architectures}
Ocean models, as part of earth system models, usually comprise multiple scientific models including one for the atmosphere and one for the ocean~\citep{collins2005design}. 
The ocean model itself includes at least a scientific model to simulate the circulation in the ocean, and a scientific model describing the behavior in the water column, e.g. sediments and plankton.
As the ocean and the atmosphere interact, the scientific models have to share information.
This is done by couplers which also map different spatial properties and resolutions, e.g., different grid and mesh sizes.
In some cases, these meshes can have varying densities.
Another option for scientific model interaction is called nesting.
Here, a global ocean model with a wide mesh is used to cover the whole globe. For specific areas, a denser mesh may be overlaid.

Architecture knowledge is often not externalized and only kept by experts.
As the development and maintenance of a model component progresses, the documentation becomes outdated.
Multiple interviewers described the quality of model documentation as `missing, outdated and wrong' and pointed out that missing is often better than the other two options.
Thus, scientists and engineers need to comprehend the model code as the only truth.

\subsection{Research done with Models}\label{ss-research-done}

Scientists use the ocean models for at least three different types of experiments:
\begin{enumerate}
	\item Scenario-based predictions, like in the assessment of climate change.
For instance, various CO\textsubscript{2} profiles could be used.
Usually, these scenarios are run multiple times to gain probability distributions.
	\item Sensitivity studies are conducted to test the ocean model's stability/sensitivity to certain changes.
These studies are used to evaluate the validity of ocean models under certain conditions and to detect tipping points.
	\item Ocean model calibration aims to find and check start values by comparing ocean model output with observation data.
\end{enumerate}

\subsection{Threats to Validity}

Our interviewees have primarily been recruited from  Kiel University, the GEOMAR Helmholtz Centre for Ocean Research Kiel, the German Climate Computing Center (DKRZ) and the Max Planck Institute for Meteorology (MPI).
Thus, our findings may not cover the complete scope of scientific model development and use.
We suggest and hope that other researchers will extend and repeat our thematic domain analysis with model developers and research software engineers working in related domains of scientific computing to further advance our knowledge and skills in scientific modeling.

\section{Conclusions and Future Work}\label{sec-conclusion}

We conducted semi-structured interviews and analyzed the interview results via Thematic Analysis~(TA) to better understand scientific software development in ocean science.
We identified needs of developers and obstacles that are a hindrance for model development and evolution.
In this paper, we discussed six themes concerning ocean modeling.
They comprise the platform and environment where ocean models are developed, how models are executed, the evolution of model code, development techniques and methods, and what kind of experiments are performed with models.
These themes help to understand where methods from software engineering could be introduced and which challenges they have to address.
The complete map is provided as supplementary material in the form of an online graph.\footnote{\url{https://oceandsl.uni-kiel.de/graph}}

While the findings in this paper are of a qualitative nature, they provide an understanding of how scientists and research software engineers work, to allow us to develop suitable methods and tools to improve their work.
Based on our analysis, we see the following areas as particularly pressing and also promising for improving ocean system modeling:
\begin{itemize}
	\item Systematic and automatic testing will help with maintaining the quality of long-living ocean models.
	\item Modular software architectures will allow for testing smaller units. 
	\item Collaborative development in scientific communities will also profit from modular software architectures, since the developers may work more independently on parts of the models.
	\item Domain-specific approaches to developing and maintaining ocean models may improve the productivity and quality of these models.
	\item Tools that help with code comprehension of existing code bases will help new developers to contribute to the models.
	\item Teaching basic software engineering skills to scientists is indispensable.
\end{itemize}
In the future, we will provide an additional analysis on the characteristics and properties of the ocean modeling domain which provides insights on social interactions, tooling and collaboration.
Based on the findings of this paper, we will develop and evaluate tooling, especially domain-specific languages \citep{ESE2017}, to improve certain aspects of working with ocean system models.

\printcredits

\bibliographystyle{cas-model2-names}

\end{document}